\documentclass[usenatbib]{mnras}
\usepackage{txfonts, color, soul,hyperref,graphicx}

\begin{document}

\title[Stellar activity cycles and magnetic field topology]{The connection between stellar activity cycles and magnetic field topology}

\author[V. See et al.]
{V. See$^1$\thanks{E-mail: wcvs@st-andrews.ac.uk}, M. Jardine$^1$, A. A. Vidotto$^{2,3}$,  J.-F. Donati$^{4,5}$, S. Boro Saikia$^{6}$,
\newauthor J. Bouvier$^{7,8}$, R. Fares$^{9}$, C. P. Folsom$^{7,8}$, S. G. Gregory$^1$, G. Hussain$^{10}$, 
\newauthor S. V. Jeffers$^{6}$, S. C. Marsden$^{11}$, J. Morin$^{12}$, C. Moutou$^{13,14}$, 
\newauthor J. D. do Nascimento Jr$^{15,16}$, P. Petit$^{4,5}$, I. A. Waite$^{11}$\\
$^{1}$SUPA, School of Physics and Astronomy, University of St Andrews, North Haugh, KY16 9SS, St Andrews, UK\\
$^{2}$Observatoire de Gen\`eve, Universit\'e de Gen\`eve, Chemin des Maillettes 51, Sauverny, CH-1290, Switzerland\\
$^{3}$School of Physics, Trinity College Dublin, University of Dublin, Dublin-2, Ireland\\
$^{4}$Universit\'{e} de Toulouse, UPS-OMP, Institut de Recherche en Astrophysique et Plan\'{e}tologie, Toulouse, France\\
$^{5}$CNRS, Institut de Recherche en Astrophysique et Plan\'{e}tologie, 14 Avenue Edouard Belin, F-31400 Toulouse, France\\
$^{6}$Universit\"at G\"ottingen, Institut f\"ur Astrophysik, Friedrich-Hund-Platz 1, 37077 G\"ottingen, Germany\\
$^{7}$Univ. Grenoble Alpes, IPAG, F-38000 Grenoble, France\\
$^{8}$CNRS, IPAG, F-38000 Grenoble, France \\
$^{9}$INAF- Osservatorio Astrofisico di Catania, Via Santa Sofia, 78 , 95123 Catania, Italy\\
$^{10}$ESO, Karl-Schwarzschild-Str. 2, D-85748 Garching, Germany\\
$^{11}$Computational Engineering and Science Research Centre, University of Southern Queensland, Toowoomba, 4350, Australia\\
$^{12}$LUPM, Universit\'e de Montpellier, CNRS, France\\
$^{13}$Canada-France-Hawaii Telescope Corporation, CNRS, 65-1238 Mamalahoa Hwy, Kamuela HI 96743, USA\\
$^{14}$Aix Marseille Universit\'{e}, CNRS, LAM (Laboratoire d'Astrophysique de Marseille) UMR 7326, 13388, Marseille, France\\
$^{15}$Departmento de F\'isica Te\'orica e Experimental, Universidade Federal do Rio Grande do Norte, CEP:59072-970 Natal, RN, Brazil\\
$^{16}$Harvard-Smithsonian Center for Astrophysics, Cambridge, Massachusetts 02138, USA}

\maketitle

\begin{abstract}
Zeeman Doppler imaging has successfully mapped the large-scale magnetic fields of stars over a large range of spectral types, rotation periods and ages. When observed over multiple epochs, some stars show polarity reversals in their global magnetic fields. On the Sun, polarity reversals are a feature of its activity cycle. In this paper, we examine the magnetic properties of stars with existing chromospherically determined cycle periods. Previous authors have suggested that cycle periods lie on multiple branches, either in the cycle period-Rossby number plane or the cycle period-rotation period plane. We find some evidence that stars along the active branch show significant average toroidal fields that exhibit large temporal variations while stars exclusively on the inactive branch remain dominantly poloidal throughout their entire cycle. This lends credence to the idea that different shear layers are in operation along each branch. There is also evidence that the short magnetic polarity switches observed on some stars are characteristic of the inactive branch while the longer chromospherically determined periods are characteristic of the active branch. This may explain the discrepancy between the magnetic and chromospheric cycle periods found on some stars. These results represent a first attempt at linking global magnetic field properties obtained form ZDI and activity cycles.
\end{abstract}


\section{Introduction}
\label{sec:Intro}
On the Sun, tracers of magnetic activity, such as sunspot number, are known to vary cyclically with a period of roughly 11 years. Analogous activity cycles are also thought to exist in other stars with outer convection zones. However, it is not possible to count starspots on unresolved stellar discs making the determination of stellar activity cycle periods a non-trivial task. One option is to measure the disc integrated emission in calcium lines as a function of time. In this regard, the Mount Wilson Observatory has played an instrumental role in advancing knowledge of stellar cycles via multi-decade chromospheric observations of solar-like stars \citep{Wilson1978,Baliunas1995,Metcalfe2013,Egeland2015}. Further observational campaigns have also been directly inspired by the work done at the Mount Wilson Observatory \citep[e.g.][]{Hall2007}. Various studies into the behaviour of chromospheric activity have resulted from these types of observations including research into chromospheric and photometric variability \citep{Lockwood2007} and the use of activity proxies as age indicators \citep{Mamajek2008,Pace2013}. Some authors have also studied possible trends involving the activity cycle duration and its relation to other stellar parameters. For example, \citet{Brandenburg1998} and \citet{Saar1999} showed that stars may lie on several branches when the ratio of their cycle frequency to the angular rotational frequency, $\omega_{\rm{cyc}}/\Omega$, is plotted against inverse Rossby number, $\rm{Ro}^{-1} = \tau_{\rm{c}}/\rm{P_{rot}}$. Here, $\tau_{\rm{c}}$ and $\rm{P_{rot}}$ are the convective turnover time and rotation period respectively. These authors called these branches the inactive, active and superactive branches. It is thought that the different branches may be a manifestation of changes in the underlying dynamos of these stars as they evolve over their lifetime. For example, \citet{BohmVitense2007} suggested that the dominant shear layer contributing to dynamo action in active branch stars is the near surface shear layer while for the inactive branch stars, it is the shear layer between the inner radiative core and outer convective zone known as the tachocline. In recent years, many authors have conducted further investigations into the nature of these branches \citep{BohmVitense2007,Arkhypov2015,Lopes2015,Lehtinen2016} as well as how activity cycles evolve over the stellar lifetime \citep{Olah2016}.

A second option for determining activity cycle periods is long term monitoring of stellar magnetic fields. On the Sun, sunspots occur as a result of emerging flux and reflect the underlying magnetic field generation mechanisms, i.e. the solar dynamo. The magnetic field topology of a star can therefore be considered a more fundamental measure of activity cycles. Indeed, the Sun's global magnetic field switches polarity roughly once every 11 years \citep{DeRosa2012}, in phase with the chromospheric activity cycle. A full magnetic cycle, i.e. two polarity switches, therefore comprises two chromospheric cycles. Additionally, theoretical dynamo simulations have also been able to reproduce polarity switches in the large-scale magnetic field of stars though the exact processes that determine the time-scale of these switches is still unclear \citep{Ghizaru2010,Brown2011,Augustson2013,Passos2014,Pipin2015}.

The monitoring of stellar magnetic field topologies can be achieved with the Zeeman-Doppler imaging (ZDI) technique. This is a tomographic technique capable of reconstructing large-scale magnetic field topologies at stellar surfaces by inverting a series of spectropolarimetric observations \citep{Donati1997}. ZDI has already been used to study magnetic trends as a function of fundamental parameters \citep{Petit2008,Donati2008,Morin2008,Morin2010,Vidotto2014,See2015}, field evolution on the pre-main sequence \citep{Gregory2012,Folsom2016} and the magnetic properties of stars with indirect mass-loss measurements \citep{Vidotto2016}. Additionally, repeated observations of individual targets have revealed that some stars undergo polarity reversals that may be indicative of activity cycles \citep{Donati2003Dynamo,Donati2008,Fares2009,Petit2009,Morgenthaler2011,Fares2013,Rosen2016,Saikia2016}. When using the global magnetic field topology as an indicator of activity cycles, we must be careful to distinguish between chromospheric cycle periods and magnetic cycle periods. In the solar context, the former has a value of $\sim$11 years while the latter has a value of $\sim$22 years \citep{DeRosa2012}. In the rest of this paper, we will refer to cycle periods determined from chromospheric activity observations as chromospheric activity cycles and cycle periods determined from magnetic field reversals as magnetic activity cycles. We must also be mindful of the fact that, due to the amount of observation time required to reconstruct a single magnetic map, the number of ZDI maps one is able to produce over an activity cycle will be much more sparse when compared to the number of chromospheric observations. Therefore, it is useful to study activity cycles with chromospheric data in conjunction with the ZDI technique.

There are now numerous stars that have been characterised by ZDI that also have a chromospherically determined cycle period in the literature. While a number of these stars have multiple ZDI maps available, many others have only been observed during one epoch. For these stars, it is clearly not possible to determine the time-scale over which they undergo polarity reversals or if reversals occur at all. However, a single ZDI map still contains useful information about the topology of the magnetic field, such as how much magnetic energy is stored in toroidal or axisymmetric modes. In this paper, we will analyse the magnetic properties of a sample of stars that have at least one ZDI map as well as a chromospheric activity cycle period determined in the literature.

\begin{figure*}
	\begin{center}
	\includegraphics[trim = 0cm 1cm 0cm 0.5cm,width=0.8\textwidth]{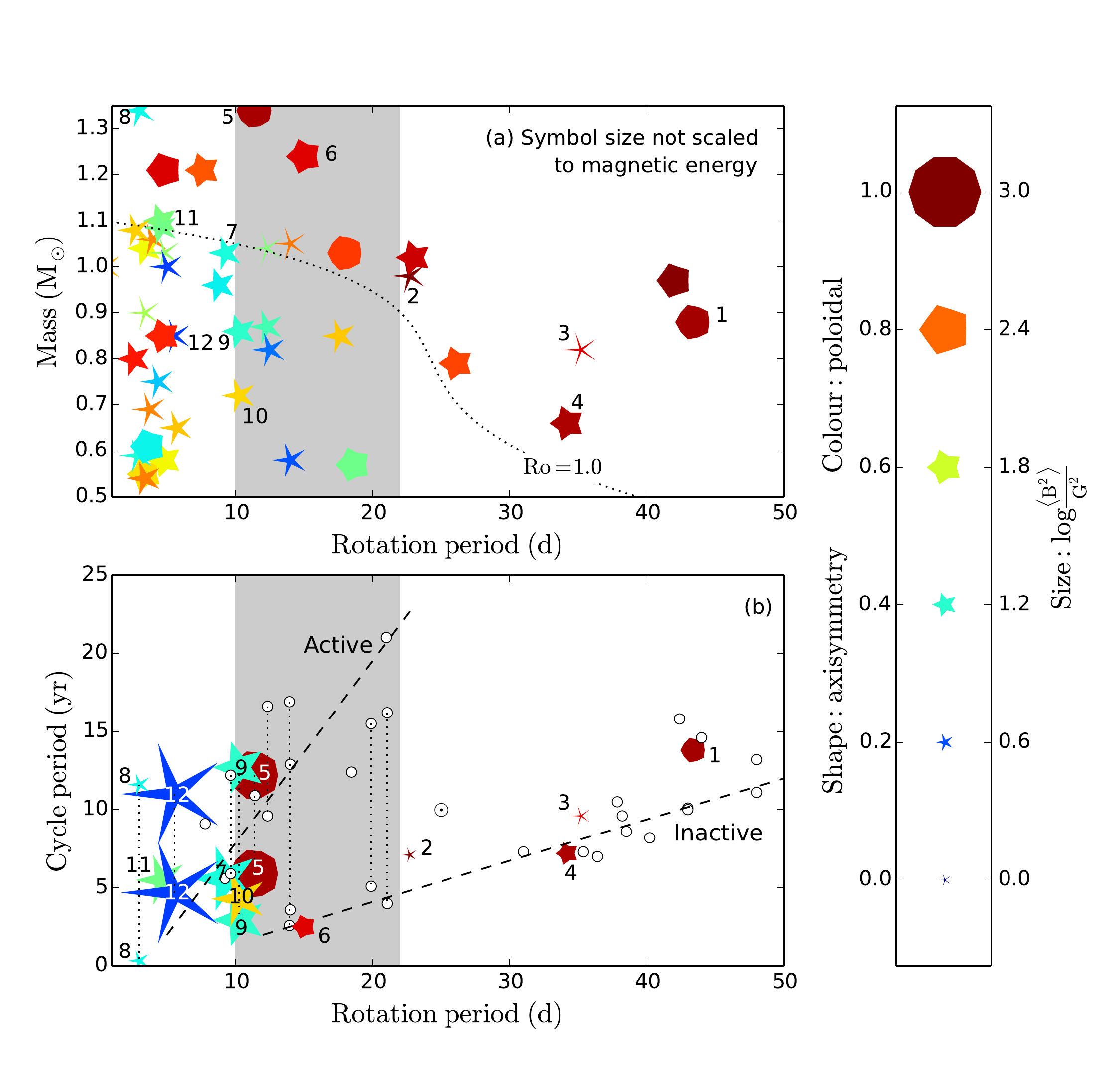}
	\end{center}
	\caption{Panel a: the sample of stars used by \citet{See2015} plotted in stellar mass-rotation period space. The stars used in this study are a subset of that sample and are labelled with a number corresponding to those found in table \ref{tab:ZDISample}. This is a similar plot to Fig. 3 of \citet{Donati2009}. The symbol colour represents the poloidal energy fraction (ranging from red for purely poloidal, i.e. $f_{\rm pol} = 1-f_{\rm tor} = 1$, to blue for purely toroidal, i.e. $f_{\rm pol} = 0$) and symbol shape represents how axisymmetric the poloidal component of the field is (ranging from decagons for a purely axisymmetric poloidal field, i.e. $f_{\rm axi,pol} = 1$, to pointed stars for a purely non-axisymmetric field, i.e. $f_{\rm axi,pol} = 0$.). Due to the large number of stars in the sample, symbol sizes have been kept the same for clarity and do not scale with $\log\langle B^2\rangle$ as is usual with this type of plot. A dotted line indicates $\rm Ro =1$. Panel b: chromospheric activity cycle period against rotation period for the sample of \citet{BohmVitense2007} plotted with open circles (see their Fig. 1). Dashed lines indicate the active and inactive branches. Overplotted is the sample outlined in section \ref{sec:Sample} where symbol colour and shape have the same meaning as panel a. In this panel, symbol size does scale with $\log \langle B^2\rangle$ as indicated by the key. Stars with multiple cycle periods are connected with a dashed line. As with panel a, each star is labelled with a number corresponding to those found in table \ref{tab:ZDISample}. On both panels the shaded region indicates the range of rotation periods where the active and inactive branches overlap.}
	\label{fig:BVConfusogram}
\end{figure*}

\begin{figure}
	\begin{center}
	\includegraphics[trim=0cm 1cm 0cm 1cm, clip = true, width=\columnwidth]{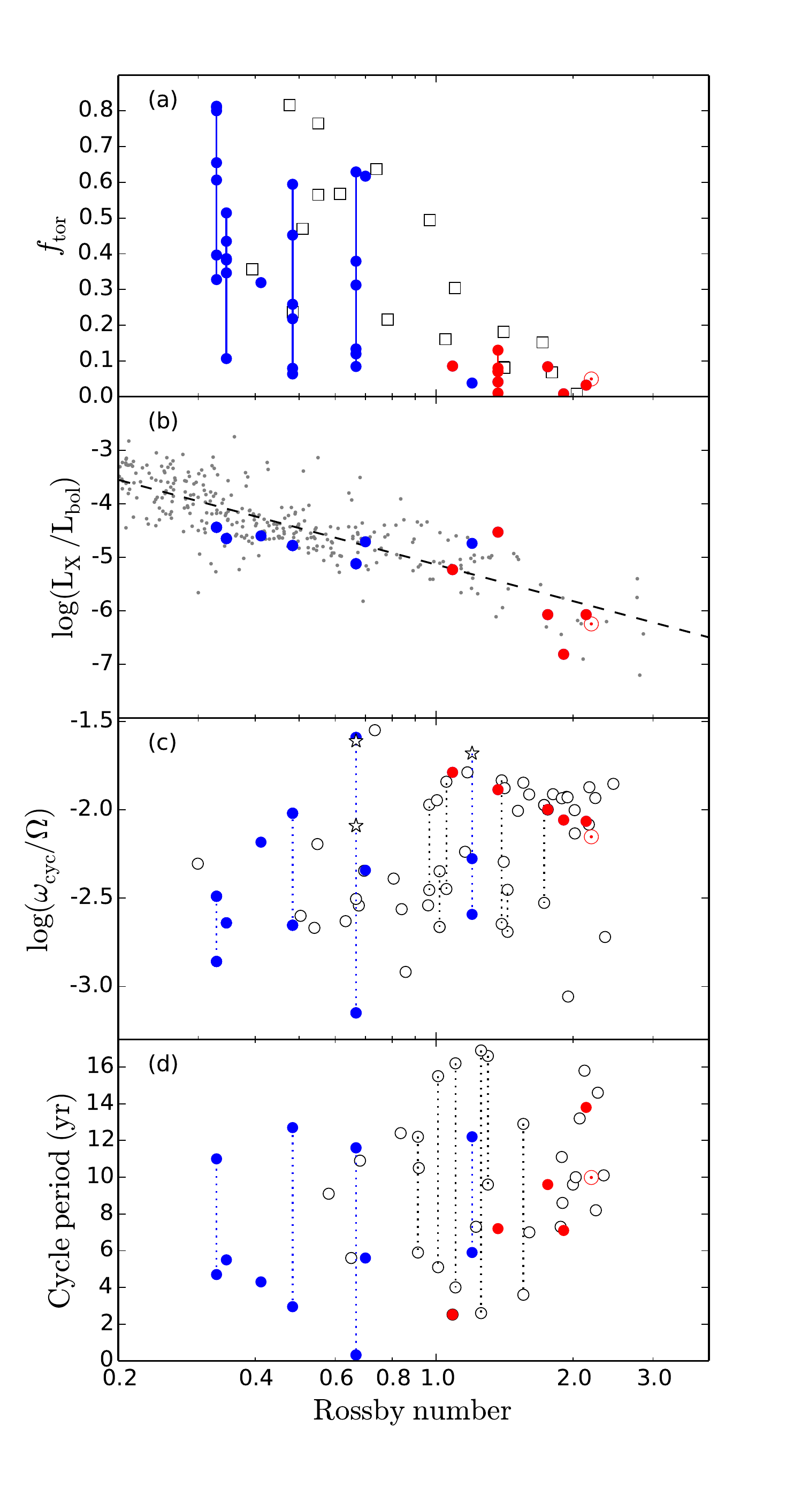}
	\end{center}
	\caption{(a) Toroidal energy fraction against Rossby number for the sample of stars used by \citet{See2015} (open square symbols). Stars observed at multiple epochs are joined by solid lines. (b) The ratio of X-ray to bolometric luminosity against Rossby number reproduced from \citet{Wright2011} (grey dots; see their Fig. 2). (c) The ratio of chromospheric cycle frequency to rotational frequency against Rossby number reproduced from \citet{Saar1999} (open circle symbols; see their Fig. 1). (d) Chromospheric cycle period against Rossby number using the sample of \citet{BohmVitense2007}. In panels c and d, stars with multiple cycles are connected by dashed lines. We also note that, in panels c and d, the quality of the chromospheric cycle period determination is better for some stars than others (see section \ref{sec:Sample}). In every panel, the Sun is shown with the solar symbol and our ZDI sample is plotted with blue and red circles denoting active and inactive branch stars respectively. In panel c, cycle periods estimated from polarity reversals for HD 78366 and $\tau$ Boo are shown with star symbols (see section \ref{subsec:MagneticCycles}). Stars are ordered by Rossby number in table \ref{tab:ZDISample} to allow for easier identification of the red and blue points in this plot.}
	\label{fig:Rossby}
\end{figure}

In section \ref{sec:Sample}, we present the sample of stars used in this study. In section \ref{sec:Results} we consider our sample within the context of previous studies. A discussion of the results and their implications is presented in section \ref{sec:Discussion} with conclusions following up in section \ref{sec:Conclusions}.

\section{Sample selection}
\label{sec:Sample}
The sample used in this study consists of stars that have both a magnetic map reconstructed using ZDI and an activity cycle period determination in the literature. Their physical parameters are listed in table \ref{tab:ZDISample}. To calculate Rossby numbers, we use the rotation periods as listed by \citet{Vidotto2014} and convective turnover times were calculated using the method described by \citet{Saar1999}. Values for the X-ray luminosity normalised to the bolometric luminosity, $\rm{R_X} = L_{\rm{X}}/L_{\rm{bol}}$, are taken from \citet{Vidotto2014} and references therein. 

The large-scale surface magnetic fields, as reconstructed from ZDI, are represented by a spherical harmonic decomposition (see \citet{Donati2006} and summary by \citet{See2015} for further details). As well as the overall magnetic field at the stellar surface, different components of the field can also be recovered by ZDI. Typically, the parameters that are of most interest are the magnetic energy density averaged over the stellar surface, $\langle B^2 \rangle$, the toroidal energy fraction, $f_{\rm tor} = \langle B_{\rm tor}^2 \rangle/\langle B^2 \rangle$ and the axisymmetric energy fraction, $f_{\rm axi} = \langle B_{\rm axi}^2 \rangle/\langle B^2 \rangle$. It is also common to look at the axisymmetric fraction of the poloidal component only, $f_{\rm axi,pol} = \langle B_{\rm axi,pol}^2 \rangle/\langle B_{\rm pol}^2 \rangle$. A large number of the stars analysed in this paper were observed as part of the BCool program (for further information on the work and goals of the BCool collaboration, see \citet{Marsden2014}). The original paper that each ZDI map is published in is listed in table \ref{tab:ZDISample}.

Since the highest order spherical harmonic order, $l_{\rm max}$, that can be reconstructed by ZDI depends on the rotation rate of the star \citep{Fares2012}, the maximum spatial resolution that can be achieved varies from star to star. It is possible that this could introduce a bias related to $l_{\rm max}$ into our results. However, the majority of the magnetic energy in ZDI reconstructions is contained in the lowest order modes \citep[e.g.][]{Petit2008,Rosen2016}. Therefore, as discussed by \citet{Vidotto2014}, the fact that different stars are reconstructed with different $l_{\rm max}$ does not significantly affect the results.

We have restricted ourselves to activity cycles periods determined from chromospheric measurements to maintain consistency across our sample (with the exception of HN Peg; see discussion at the end of this section). For example, chromospherically determined cycle periods can differ from those determined from photometry for a number of reasons \citep[e.g.][]{Messina2002}. A large number of the chromospheric cycle determinations come from \citet{Baliunas1995} though some come from other sources, the references for which are listed in table \ref{tab:ZDISample}. We have included a number of cycle periods that have been classified as `poor' or `fair' by \citet{Baliunas1995} under their false alarm probability (FAP) grading scheme in order to improve the number of objects in this study (these are noted in table \ref{tab:ZDISample}). Accordingly, when interpreting the results, these objects may need to be treated more cautiously. However, they do not seem to be discrepant with the rest of the sample and our conclusions are not dependent on these less reliable chromospheric cycle periods. We discuss some individual cases here:

\textbf{61 Cyg A}: This star is a K dwarf with a well known chromospheric activity cycle of approximately 7 years \citep{Baliunas1995}.  From observations taken at the NARVAL spectropolarimeter on the Telescope Bernard Lyot, together with old archival data, \citet{Saikia2016} determine a chromospheric cycle of 7.2$\pm$1.3 years. This value is in agreement with the long cycle period determined by \citet{Olah2009} from Ca II data. \citet{Olah2009} also found a secondary chromospheric period of 3.6 years in part of their data. However, \citet{Saikia2016} find no evidence of this shorter period and so we will only use the 7.2 year period. Additionally, 61 Cyg A exhibits an X-ray activity cycle which is in phase with the chromospheric activity cycle \citep{Robrade2012}.  Long term spectropolarimetric monitoring of this star has also revealed a solar-like magnetic cycle \citep{Saikia2016}, which makes it the first cool star other than the Sun where the magnetic and chromospheric activity cycles are in phase.

\textbf{$\tau$ Boo}: \citet{Baliunas1995} found a 11.6 year period but assign it a poor grade in their false alarm probably classification scheme calling into question the reliability of this period determination. However, we still include this object in our sample since it is interesting in the context of magnetic activity cycles (see section \ref{subsec:MagneticCycles}). Additionally \citet{Baliunas1997} and \citet{Mengel2016} both report a chromospheric cycle period of around 116 days. 

\textbf{HN Peg}: This star was also assigned a poor grade by \citet{Baliunas1995} who found a period of 6.2 years. \citet{Messina2002} found a 5.5 year period from an analysis based on photometric data and we use this value due to the smaller false alarm probability that these authors find. However, both values are compatible with the active branch of stars and so our results are unaffected by the choice of one value over the other.

\textbf{HD 78366}: \citet{Baliunas1995} reported cycle periods of 12.2 years (good FAP grade) and 5.9 years (fair FAP grade). Recently, Jeffers et al. (in prep) reconstructed the field of HD 78366 over four observational epochs. These authors found that the variation in the star's S-index over these four epochs are not inconsistent with the 5.9 year period of \citet{Baliunas1995}. We will use both the 12.2 year and 5.9 year cycle in the rest of this work but it is worth being cautious with this particular cycle period given its fair grade. We note that our conclusions are not dependent on the 5.9 year cycle.

\section{Results}
\label{sec:Results}
\subsection{Field properties}
\label{subsec:FieldProperties}
In this section we discuss the magnetic properties of our sample. Due to the relatively small number of stars in our sample, we will also draw on other studies with larger sample sizes. We discuss these magnetic trends in relation to previous work on magnetic activity and activity cycles.

In Fig. \ref{fig:BVConfusogram}a, we plot the magnetic properties of the sample of stars used by \citet{See2015} in stellar mass-rotation period space similarly to Fig. 3 of \citet{Donati2009}. The symbol colour scales with the poloidal energy fraction, $f_{\rm pol}= 1-f_{\rm tor}$, and the symbol shape scales with the axisymmetry of the poloidal component, $f_{\rm axi,pol}$. Numerous authors have used this method of representing magnetic field characteristics in various different parameter spaces \citep{Donati2008,Morin2008,Donati2009,Morin2010,Vidotto2016,Folsom2016}. Usually, the symbol size scales with $\log\langle B^2\rangle$ on this type of plot. However, due to the density of points in this plot, we have chosen not to do so here for clarity. For stars with multiple ZDI maps, we have only plotted the epoch with the largest $f_{\rm tor}$ value as this highlights the difference between stars that are always dominantly poloidal and those that show large fluctuations in their toroidal energy fractions. Additionally, we have restricted the parameter space to stars more massive than 0.5$\rm M_{\odot}$ since less massive stars likely have different dynamo mechanisms to the stars we analyse in this study \citep{Donati2008,Morin2008,Morin2010,Gregory2012,Yadav2015}. As outlined by \citet{Donati2009}, the Rossby number is important in the context of magnetic field topologies. A clear transition at a Rossby number of ${\sim}1$ (dotted line) can be seen in the field topologies. Stars with $\rm Ro \gtrsim 1$ (top right of plot) mostly show dominantly poloidal and axisymmetric fields whereas $\rm Ro \lesssim 1$ stars (bottom left of plot) are capable of generating significant or even dominantly toroidal fields that are non-axisymmetric.

In the context of stellar activity, the preference of Rossby number over rotation period is motivated from both empirical \citep{Wright2011} and theoretical considerations \citep{Noyes1984}. However, it is worth noting that some authors have argued that rotation period is the more fundamental parameter \citep{Reiners2014}. In Fig. \ref{fig:Rossby}a, we plot the toroidal energy fraction, $f_{\rm tor}$, directly against Rossby number with red and blue circles (these colours correspond to inactive and active branch stars respectively; see section \ref{subsec:Branches}). Stars that have been observed at multiple epochs are connected by solid lines. Additionally, we plot the stars in the sample used by \citet{See2015} with open square markers and the Sun during Carrington Rotation CR2109 (shortly after solar minimum) with a solar symbol\footnote{The solar value of $f_{\rm tor}$ used in Fig. \ref{fig:Rossby}a ($f_{\rm tor}=0.05$) is obtained for a synoptic map truncated to $l_{\rm max}=5$. As discussed by \citet{Vidotto2016}, this provides a fairer comparison to ZDI maps that only capture the large scale field structures. We note that our choice of $l_{\rm max}$ does not drastically affect the toroidal energy fraction. Indeed $f_{\rm tor}$ remains below 0.1 for any choice of $l_{\rm max}$ (see Fig. 5 of \citet{Vidotto2016}).} \citep{Vidotto2016Sun}. This plot is similar to Fig. 6 of \citet{Petit2008}. These authors studied four stars that were roughly one solar mass each and showed that the poloidal energy fraction, $f_{\rm pol}$, increases with rotation period. We see a similar behaviour here whereby the stars with the longest rotation periods (largest $\rm Ro$) display dominantly poloidal fields. Conversely, the most rapidly rotating stars (smallest $\rm Ro$) show large $f_{\rm tor}$ variations and are capable of developing dominantly toroidal fields. As in Fig. \ref{fig:BVConfusogram}a, the transition between these two regimes occurs at $\rm{Ro} \sim 1.0$. This behaviour has also been noted by \citet{Donati2009} and \citet{Folsom2016}. Comparing with the expanded sample in Fig. \ref{fig:Rossby}a, we see that the four stars of \citet{Petit2008} trace the upper envelope of points shown here. On a plot of $f_{\rm tor}$ against rotation period (not shown), we find that the transition from dominantly poloidal stars to stars that are able to generate dominantly toroidal fields occurs at a rotation of $\sim$12 days. This is in agreement with Fig. \ref{fig:BVConfusogram}b and is consistent with the analysis of \citet{Petit2008}. We also find that this rotation period separates the inactive and active branch stars in our sample.

\begin{table*}
\begin{minipage}{160mm}
	\caption{Parameters for the stars used in this study, ordered by Rossby number: label used to identify each star in Fig. \ref{fig:BVConfusogram}, spectral type, mass, rotation period, convective turnover time, Rossby number, primary and secondary cycle period (if one exists), X-ray to bolometric luminosity ratio, toroidal energy fraction and the epoch of the observations from which each ZDI map was reconstructed. Each star is categorised as an active (A) or inactive (I) branch star corresponding to the blue and red points in Fig. \ref{fig:Rossby}. The paper from which cycle periods are taken are referenced with a superscript on each cycle period value. Similarly, the paper where each magnetic map was originally published is referenced with a superscript in the observation epoch column. Cycle periods listed as fair or poor under the false alarm probability scheme of \citet{Baliunas1995} are shown in brackets. Convective turnover times are from \citet{Saar1999}. For the remaining parameters, references can be found in \citet{Vidotto2014}. } 
	\label{tab:ZDISample}
	\begin{tabular}{lcccccccccccc}
		\hline
		Star & Figure & Spec. & $M_{\star}$	& $\rm{P}_{\rm{rot}}$	& $\tau_{\rm{c}}$ & $\rm{Ro}$	& $\rm{P}_{\rm{cyc}}$	&	$\rm{P}_{\rm{cyc,2}}$	&	$\log \rm L_{X}/L_{bol}$ & $f_{\rm{tor}}$ &	ZDI obs & Branch\\
		ID & Label & Type & [$M_{\odot}$] & [d] & [d] & & [yr] & [yr] & & & epoch &\\
		\hline
Sun	&	-	&	G2V	&	1	&	26.09	&	11.9	&	2.19	&	10\textsuperscript{(1)}	&	-	&	-6.24	&	0.05	&	2011 Apr\textsuperscript{(2)}	&	I	\\
HD 3651	&	1	&	K0V	&	0.88	&	43.4	&	20.3	&	2.14	&	13.8\textsuperscript{(1)}	&	-	&	-6.07	&	0.03	&	-\textsuperscript{(3)}	&	I	\\
18 Sco	&	2	&	G2V	&	0.98	&	22.7	&	11.9	&	1.91	&	7.1\textsuperscript{(4)}	&	-	&	-6.81	&	0.01	&	2007 Aug\textsuperscript{(5)}	&	I	\\
HD 10476	&	3	&	K1V	&	0.82	&	35.2	&	20	&	1.76	&	9.6\textsuperscript{(1)}	&	-	&	-6.07	&	0.08	&	-\textsuperscript{(3)}	&	I	\\
61 Cyg A	&	4	&	K5V	&	0.66	&	34.2	&	25	&	1.37	&	7.2\textsuperscript{(6)}	&	-	&	-4.53	&	0.04	&	-\textsuperscript{(3)}	&	I	\\
...	&	...	&	...	&	...	&	...	&	...	&	...	&	...	&	...	&	...	&	0.07	&	2007 Jul\textsuperscript{(6)}	&	...	\\
...	&	...	&	...	&	...	&	...	&	...	&	...	&	...	&	...	&	...	&	0.08	&	2008 Aug\textsuperscript{(6)}	&	...	\\
...	&	...	&	...	&	...	&	...	&	...	&	...	&	...	&	...	&	...	&	0.13	&	2010 Jun\textsuperscript{(6)}	&	...	\\
...	&	...	&	...	&	...	&	...	&	...	&	...	&	...	&	...	&	...	&	0.01	&	2013 Jul\textsuperscript{(6)}	&	...	\\
...	&	...	&	...	&	...	&	...	&	...	&	...	&	...	&	...	&	...	&	0.07	&	2014 Jul\textsuperscript{(6)}	&	...	\\
...	&	...	&	...	&	...	&	...	&	...	&	...	&	...	&	...	&	...	&	0.13	&	2015 Jun\textsuperscript{(6)}	&	...	\\
HD 78366	&	5	&	F9V	&	1.34	&	11.4	&	9.5	&	1.20	&	12.2\textsuperscript{(1)}	&	(5.9)\textsuperscript{(1)}	&	-4.74	&	0.04	&	-\textsuperscript{(3)}	&	A	\\
HD 76151	&	6	&	G3V	&	1.24	&	15	&	13.8	&	1.09	&	(2.52)\textsuperscript{(1)}	&	-	&	-5.23	&	0.09	&	2007 Feb\textsuperscript{(5)}	&	I	\\
$\kappa$ Ceti	&	7	&	G5V	&	1.03	&	9.3	&	13.3	&	0.70	&	(5.6)\textsuperscript{(1)}	&	-	&	-4.71	&	0.62	&	2012 Oct\textsuperscript{(7)}	&	A	\\
$\tau$ Boo	&	8	&	F7V	&	1.34	&	3	&	4.5	&	0.67	&	(11.6)\textsuperscript{(1)}	&	0.32\textsuperscript{(8)}	&	-5.12	&	0.63	&	2008 Jan\textsuperscript{(9)}	&	A	\\
...	&	...	&	...	&	...	&	...	&	...	&	...	&	...	&	...	&	...	&	0.08	&	2008 Jun\textsuperscript{(9)}	&	...	\\
...	&	...	&	...	&	...	&	...	&	...	&	...	&	...	&	...	&	...	&	0.13	&	2008 Jul\textsuperscript{(9)}	&	...	\\
...	&	...	&	...	&	...	&	...	&	...	&	...	&	...	&	...	&	...	&	0.12	&	2009 May\textsuperscript{(10)}	&	...	\\
...	&	...	&	...	&	...	&	...	&	...	&	...	&	...	&	...	&	...	&	0.38	&	2010 Jan\textsuperscript{(10)}	&	...	\\
...	&	...	&	...	&	...	&	...	&	...	&	...	&	...	&	...	&	...	&	0.31	&	2011 Jan\textsuperscript{(10)}	&	...	\\
$\epsilon$ Eri	&	9	&	K2V	&	0.86	&	10.3	&	21.3	&	0.48	&	2.95\textsuperscript{(11)}	&	12.7\textsuperscript{(11)}	&	-4.78	&	0.08	&	2007 Jan\textsuperscript{(12)}	&	A	\\
...	&	...	&	...	&	...	&	...	&	...	&	...	&	...	&	...	&	...	&	0.06	&	2008 Jan\textsuperscript{(12)}	&	...	\\
...	&	...	&	...	&	...	&	...	&	...	&	...	&	...	&	...	&	...	&	0.59	&	2010 Jan\textsuperscript{(12)}	&	...	\\
...	&	...	&	...	&	...	&	...	&	...	&	...	&	...	&	...	&	...	&	0.26	&	2011 Oct\textsuperscript{(12)}	&	...	\\
...	&	...	&	...	&	...	&	...	&	...	&	...	&	...	&	...	&	...	&	0.45	&	2012 Oct\textsuperscript{(12)}	&	...	\\
...	&	...	&	...	&	...	&	...	&	...	&	...	&	...	&	...	&	...	&	0.22	&	2013 Sep\textsuperscript{(12)}	&	...	\\
$\xi$ Boo B	&	10	&	K4V	&	0.72	&	10.3	&	25	&	0.41	&	4.3\textsuperscript{(13)}	&	-	&	-4.6	&	0.32	&	-\textsuperscript{(3)}	&	A	\\
HN Peg	&	11	&	G0V	&	1.1	&	4.55	&	13.3	&	0.34	&	5.5\textsuperscript{(14)}	&	-	&	-4.65	&	0.5	&	-\textsuperscript{(3)}	&	A	\\
...	&	...	&	...	&	...	&	...	&	...	&	...	&	...	&	...	&	...	&	0.51	&	2008 Aug\textsuperscript{(15)}	&	...	\\
...	&	...	&	...	&	...	&	...	&	...	&	...	&	...	&	...	&	...	&	0.11	&	2009 Jun\textsuperscript{(15)}	&	...	\\
...	&	...	&	...	&	...	&	...	&	...	&	...	&	...	&	...	&	...	&	0.35	&	2010 Jul\textsuperscript{(15)}	&	...	\\
...	&	...	&	...	&	...	&	...	&	...	&	...	&	...	&	...	&	...	&	0.39	&	2011 Jul\textsuperscript{(15)}	&	...	\\
...	&	...	&	...	&	...	&	...	&	...	&	...	&	...	&	...	&	...	&	0.38	&	2013 Jul\textsuperscript{(15)}	&	...	\\
$\xi$ Boo A	&	12	&	G8V	&	0.85	&	5.56	&	16.9	&	0.33	&	4.7\textsuperscript{(13)}	&	11\textsuperscript{(13)}	&	-4.44	&	0.81	&	-\textsuperscript{(3)}	&	A	\\
...	&	...	&	...	&	...	&	...	&	...	&	...	&	...	&	...	&	...	&	0.4	&	2008 Feb\textsuperscript{(3)}	&	...	\\
...	&	...	&	...	&	...	&	...	&	...	&	...	&	...	&	...	&	...	&	0.61	&	2009 July\textsuperscript{(3)}	&	...	\\
...	&	...	&	...	&	...	&	...	&	...	&	...	&	...	&	...	&	...	&	0.66	&	2010 Jan\textsuperscript{(3)}	&	...	\\
...	&	...	&	...	&	...	&	...	&	...	&	...	&	...	&	...	&	...	&	0.33	&	-\textsuperscript{(3)}	&	...	\\
...	&	...	&	...	&	...	&	...	&	...	&	...	&	...	&	...	&	...	&	0.81	&	2010 Aug\textsuperscript{(3)}	&	...	\\
...	&	...	&	...	&	...	&	...	&	...	&	...	&	...	&	...	&	...	&	0.8	&	2011 Feb\textsuperscript{(3)}	&	...	\\
	\hline
	\end{tabular}
$^{(1)}$: \citet{Baliunas1995}; $^{(2)}$: \citet{Vidotto2016Sun}; $^{(3)}$: Petit et al. (in prep); $^{(4)}$: \citet{Hall2007}; $^{(5)}$: \citet{Petit2008}; $^{(6)}$: \citet{Saikia2016}; $^{(7)}$: \citet{Nascimento2014}; $^{(8)}$: \citet{Baliunas1997}; $^{(9)}$: \citet{Fares2009}; $^{(10)}$: \citet{Fares2013}; $^{(11)}$: \citet{Metcalfe2013}; $^{(12)}$: \citet{Jeffers2014}; $^{(13)}$: \citet{Olah2009}; $^{(14)}$: \citet{Messina2002}; $^{(15)}$: \citet{Saikia2015}
	\end{minipage}
\end{table*}

\subsection{Activity-rotation relation}
\label{subsec:ActRotRel}
Coronal X-ray emission is a reliable indicator of stellar magnetic activity. Other than heating from magnetic sources, there are few plausible mechanisms that can easily induce it. The relationship between the ratio of X-ray to bolometric luminosity, $\rm R_X = L_{X}/L_{bol}$ and Rossby number is known as the activity-rotation relation and is well studied \citep{Noyes1984,Pizzolato2003,Wright2011}. In the so-called unsaturated regime, stars show increasing $\rm L_{X}/L_{bol}$ values with decreasing Rossby number until a critical Rossby number of $\rm Ro \sim 0.1$. At smaller Rossby numbers, in the so called saturated regime, X-ray emissions saturate at roughly $ \rm{L_{X}/L_{bol}}\sim 10^{-3}$. Recent studies have shown that the energy stored in large-scale magnetic fields also display the same behaviour as $\rm L_{X}/L_{bol}$, separating into the saturated and unsaturated regimes \citep{Vidotto2014,See2015,Folsom2016}. In Fig. \ref{fig:Rossby}b, we plot $\rm L_{X}/L_{bol}$ against Rossby number for both our sample (red and blue markers) and the sample of \citet{Wright2011} (grey dots) for context. It is worth noting that \citet{Wright2011} and \citet{Saar1999} use different methods to derive their convective turnover times. This might result in a small systematic difference between our sample and the sample of \citet{Wright2011}. Figure \ref{fig:Rossby}b clearly shows that our sample lies in the unsaturated regime of the activity-rotation relation. It is interesting to note that the activity-rotation relation is continuous at $\rm Ro \sim 1.0$ while there appears to be a segregation of activity branches (red and blue points) at this Rossby number. Given that activity cycles and coronal X-ray emission are both a result of dynamo activity, this is perhaps surprising.

\subsection{Activity cycle branches}
\label{subsec:Branches}
Many studies have examined the possibility that activity cycle periods may lie on multiple branches. \citet{Brandenburg1998} and \citet{Saar1999} investigated this phenomenon in the $\omega_{\rm cyc}/\Omega$ vs $\rm Ro^{-1}$ parameter space\footnote{\citet{Saar1999} use an alternative Rossby number definition to the one given here; $\rm{Ro_{SB}} = P_{\rm{rot}}/4\rm{\pi}\tau_{\rm{c}} = \rm{Ro}/4\rm{\pi}$. In this paper, we will use the definition outlined in the main body of text, i.e. $\rm P_{rot}/\tau_{\rm c}$, and convert values quoted by \citet{Saar1999} to this definition when necessary.}. These authors suggested that a given star can lie on one of two branches, or on both if it has two cycle periods, and labelled these branches as `active' or `inactive'. In Fig. \ref{fig:Rossby}c, we reproduce Fig. 1 of \citet{Saar1999} with open circle markers. Stars with two cycle period determinations are joined with a dashed line. We note that our plot is reversed compared with the plot of \citet{Saar1999} because we plot against Rossby number rather than inverse Rossby number. Additionally, there is a range of reliability in the cycle period values used by these authors (the reliability of the cycle periods is extensively discussed in their section 2.1). We over plot our sample of stars using red and blue circles to represent stars on the inactive and active branches respectively. This colour scheme is also used for Figs. \ref{fig:Rossby}a and \ref{fig:Rossby}b. The decision of which branch a given star is assigned to is made by eye based on their position in $\omega_{\rm cyc}/\Omega$ vs $\rm Ro$ parameter space. We have coloured a star blue if it appears to have cycles on both branches. \citet{Saar1999} also discuss the possibility of a third branch of very rapid rotators ($\rm Ro\lesssim 0.1$). Since our sample lacks $\rm{Ro}\lesssim 0.1$ stars, we will not consider this branch in our analysis.

\citet{BohmVitense2007} also considered the possibility that activity cycle periods may lie on multiple branches. This author studied the stars from \citet{Baliunas1995} with the most reliable chromospheric cycle period determinations. We reproduce their plot of cycle period against rotation period with open circles in Fig. \ref{fig:BVConfusogram}b (c.f. with Fig. 1 of \citet{BohmVitense2007}) with our sample overplotted. The symbol colour and shape for our sample have the same format as Fig. \ref{fig:BVConfusogram}a. Additionally the symbol sizes scale with $\log \langle B^2\rangle$ unlike in Fig. \ref{fig:BVConfusogram}a. Interestingly, the Sun appears to be an outlier in this parameter space since it does not lie on either branch. However, when plotted in $\omega_{\rm cyc}/\Omega$ vs $\rm Ro$ space, the Sun clearly lies on the inactive branch (see Fig. \ref{fig:Rossby}c).

\citet{BohmVitense2007} deliberately chose to avoid special numbers from dynamo theory, including the Rossby number in her study. However, given the importance of this parameter to magnetic topologies and activity, we also wanted to investigate how it affects activity cycle periods. Fig. \ref{fig:Rossby}d shows the sample of \citet{BohmVitense2007} plotted in activity cycle period-Rossby number space (open circles). Additionally our sample is also plotted in red and blue circles. These colours have the same meaning as in the rest of Fig. \ref{fig:Rossby}. The inactive branch can be seen as a sequence extending down the right hand side of the plot (most easily seen by following the red points). The active branch is less obvious but can still be seen in this plot.

\section{Discussion}
\label{sec:Discussion}
\subsection{Large-scale field geometry along activity branches}
\label{subsec:BranchGeometry}
Fig. \ref{fig:BVConfusogram}b shows that all the inactive branch stars are strongly poloidal while the active branch stars can have strong toroidal fields. This is also evident from Fig. \ref{fig:Rossby}a where the inactive branch stars (red points) are all dominantly poloidal while the active branch stars (blue points) show large $f_{\rm tor}$ variations. We therefore propose the hypothesis that stars on the two branches have distinct magnetic field topologies - dominantly poloidal fields on the inactive branch while active branch stars display significant toroidal fields with large temporal variations in the toroidal energy fraction. We will discuss a potential problem with this hypothesis caused by an idiosyncrasy in our sample in section \ref{subsec:True?}. Before moving on, it is worth discussing the active branch star, HD 78366. In Fig. \ref{fig:BVConfusogram}b (labelled 5), it looks as if it might be discrepant due to its strongly poloidal fields. However, this star has not been observed over its full activity cycle. Without further observations, it is not possible to tell whether it is truly discrepant or whether it was just coincidentally observed during a part of its cycle when it was in a poloidal state. It is also worth noting that HD 78366 has a relatively high Rossby number despite its short rotation period due to its early spectral type. 

Given that there are only five inactive stars, four of which have only been observed during one epoch each, one might question whether these stars would display large $f_{\rm tor}$ variations over a cycle. However, 61 Cyg A has been observed at six epochs over the course of its seven year cycle \citep{Saikia2016}. These authors showed that this star remained almost entirely poloidal throughout their observations. This suggests that inactive branch stars remain largely poloidal even after considering activity cycle variations.

An explanation for the differing magnetic topologies on each branch may lie in the dynamos of these stars. It is thought that strong shearing, i.e. an $\Omega$ effect, can generate toroidal field from a poloidal field (though this is not the only manner in which toroidal fields can be generated). \citet{BohmVitense2007} propose that the dominant shear layer for inactive branch stars is the interface between the radiative core and the outer convective layer, i.e. the tachocline, while for active branch stars, the dominant shear layer is the near surface shear layers. For stars with periods on both branches, both shear layers would contribute significantly. Since the tachocline lies at a greater fractional depth, flux generated there takes longer to rise and emerge at the stellar surface. In contrast, flux generated in near surface shear layers takes less time emerge and may be more likely to emerge in a stressed or toroidal state. This may explain why it is only the active branch stars that are able to possess dominantly toroidal fields. Under this interpretation, one would expect stars with cycle periods on both branches to display large $f_{\rm{tor}}$ variations throughout their cycles since the near surface shear layer and the tachocline would both be contributing to dynamo action. This is the behaviour shown by $\epsilon$~Eri and $\tau$ Boo, which are the only stars that we have ZDI maps for that have cycle periods on both branches. However, \citet{Broomhall2012} find some evidence that the short quasi-biennial variations of the sun may originate in the near surface shear layers. This appears to be a contradiction to the suggestion that the dominant shear layer for short cycle period (inactive branch) stars is the tachocline while for long cycle period (active branch) stars, it is the near surface shear layers. \citet{Metcalfe2013} speculates that the rotational history of the Sun makes it an outlier while the preliminary analysis of \citet{Nascimento2015} suggests that the Sun might be part of a previously unrecognised branch.

We can also gain further insight from the observations by comparing Figs. \ref{fig:BVConfusogram}a and \ref{fig:BVConfusogram}b. These figures are split into three regions as indicated by the shaded background. To the left and right of the shaded region, we find only active and inactive branch stars respectively corresponding roughly to $\rm P_{rot} \lesssim 10$ days and $\rm P_{rot} \gtrsim 22$ days. Within the shaded region, the active and inactive branches overlap. Looking at Fig. \ref{fig:BVConfusogram}a, we see that the shape of the $\rm Ro = 1$ curve in stellar mass-rotation period space dictates the magnetic geometry along each of the branches. To the right of the shaded region, most of the stars have $\rm Ro \gtrsim 1$ and, hence, are dominantly poloidal explaining  why we find poloidal stars on the inactive branch. Conversely, to the left of the shaded region, most of the stars have $\rm Ro \lesssim 1$ and, hence, are capable of generating strong toroidal fields explaining the toroidal stars we find on the inactive branch. In the intermediate region, we find a mix of $\rm Ro \gtrsim 1$ and $\rm Ro \lesssim 1$ stars and, hence, a mix of poloidal and toroidal stars. These may correspond to stars on the inactive and active branches respectively though currently, it is not possible to tell due to the very small number of stars with both a ZDI map and a chromospheric activity cycle period determination in this intermediate region.

\subsection{Magnetic vs chromospheric cycles}
\label{subsec:MagneticCycles}
Long term ZDI observations have shown that stellar magnetic fields are inherently variable \citep[e.g.][]{Donati2003Dynamo,Petit2009}. Of particular interest are stars that show polarity reversals analogous to the $\sim$22 year magnetic cycle of the Sun. Based on two polarity reversals, \citet{Morgenthaler2011} suggested that HD 78366 could have a magnetic cycle of $\sim$3 years while several authors have studied $\tau$ Boo determining that the most probably value for its magnetic cycle period is 2 years or 8 months \citep{Donati2008,Fares2009,Fares2013,Mengel2016}. \citet{Poppenhaeger2012} were unable to find indications of this short activity cycle in X-ray observations of $\tau$ Boo though this may be due to the sparse sampling of their data or the fact that X-ray cycles can be difficult to detect \citep{McIvor2005}. Three dimensional magnetohydrodynamic simulations of $\tau$ Boo also suggest that the X-ray cycle would be difficult to detect \citep{Vidotto2012,Nicholson2016}.

The short magnetic cycle of HD 78366 appears to be at odds with the much longer cycle period determined from chromospheric activity observations \citep{Baliunas1995}. However, there may be no discrepancy between the two sets of values. In the solar case, the chromospheric cycle period is half the length of the magnetic cycle period. If we assume that this is also the case for the short magnetic cycle period of HD 78366, we can predict $\log \omega_{\rm{cyc}} / \Omega = \log \rm{P_{rot}/P_{cyc}} = \log \frac{11.4 \rm days}{1.5 \rm years} = -1.68$. We plot this value with a star symbol in Fig. \ref{fig:Rossby}c and see that it roughly coincides with the inactive branch. It seems that the magnetic cycle period determined from ZDI for HD 78366 may be characteristic of the inactive branch while the chromospherically determined period is characteristic of the active branch. If this is true, then one would expect chromospheric observations with a time sampling of sufficient density to find an additional chromospheric cycle period of roughly 1.5 years for HD 78366. The data for $\tau$ Boo, which has a similar spectral type to HD 78366, would also seem to favour such an interpretation. Just like HD 78366, $\tau$ Boo also has a long chromospheric cycle \citep[11.6 years;][]{Baliunas1995} and a short magnetic cycle (2 years or 8 months). However, in this case a shorter chromospheric cycle that is associated with the magnetic cycle has also been detected \citep[116 days;][]{Baliunas1997,Mengel2016}. Similarly to HD 78366, we predict a cycle length for $\tau$ Boo from the two most likely time-scales (2 years or 8 months) for the magnetic polarity flips and plot these with stars on Fig. \ref{fig:Rossby}c. If this scenario is true, HD 78366 finds itself in a curious position of having three cycle periods (two chromospherically determined cycles \citep{Baliunas1995} and a short magnetic cycle \citep{Morgenthaler2011}) that cannot be explained by two dynamo modes as \citet{BohmVitense2007} suggests. We do note that the shorter chromospherically determined cycle period is only assigned a false alarm probability of `fair'  by \citet{Baliunas1995}. Additional, we also note that \citet{Baliunas1995} assigned a false alarm probability of `poor' to the 11.6 year chromospheric cycle period that they determined for $\tau$ Boo. Correspondingly, the discussion in this section should be treated with caution. 

Currently, there are very few stars on which regular polarity reversals have been observed. Looking at the sample of \citet{Saar1999}, HD 190406 has a relatively short chromospheric cycle period (2.6 years). If the magnetic fields of this star does undergo regular polarity reversals, its relatively short period makes it an attractive target.

\begin{table}
\begin{center}
\begin{minipage}{87mm}
	\caption{Our results suggest that inactive branch stars are dominantly poloidal while active branch stars are able to generate strong toroidal fields. However, it is currently unclear if this result is due to a degeneracy in our sample (see section \ref{subsec:True?} for further discussion). In this table, we present a list of ZDI targets that would help break the degeneracy in the sample. For each star, the stellar mass, rotation period, primary and secondary cycle period (if one exists), Rossby number, apparent magnitude and average S index are listed. Unless noted below, stellar masses are obtained from \citet{Takeda2007}, rotation periods, cycle periods and magnitudes from \citet{BohmVitense2007}, convective turnover times (to calculate Rossby numbers) from \citet{Saar1999} and average S index from \citet{Baliunas1995}. Additionally, these stars would fill in the gap at $\rm Ro\sim1$ in Fig. \ref{fig:Rossby}a as stars transition from dominantly poloidal to being able to generate significant toroidal energy fractions.}
	\label{tab:ZDITargets}
	\begin{tabular}{lccccccc}
		\hline
		Star & $M_{\star}$ & $\rm P_{rot}$ & $\rm P_{cyc}$ & $\rm P_{cyc,2}$ & $\rm Ro$ & $m_{\rm V}$ & $\langle S \rangle$\\
		ID & $M_{\odot}$ & [d] & [yr] & [yr] & & & \\
		\hline
HD 114710	&	1.147	&	12.35	&	16.6	&	9.6	&	1.44	&	4.26	&	0.201	\\
HD 190406	&	1.069	&	13.94	&	16.9	&	2.6	&	1.39	&	5.8	&	0.194	\\
HD 115404	&	0.86$\rm ^{(a)}$	&	18.47	&	12.4	&	-	&	0.81	&	6.52	&	0.535	\\
HD 149661	&	0.892	&	21.07	&	16.2	&	4	&	1.05	&	5.75	&	0.339	\\
HD 165341	&	0.89$\rm ^{(b)}$	&	19.9	&	15.5	&	5.1	&	0.97	&	4.03	&	0.392	\\
	\hline
\end{tabular}
$\rm ^{(a)}$: \citet{Marsden2014}, $\rm ^{(b)}$: \citet{Fernandes1998}
\end{minipage}
\end{center}
\end{table}

\subsection{Breaking the degeneracy in rotation period/Rossby number}
\label{subsec:True?}
The sample of stars with measured magnetic field geometries and chromospheric activity cycles is currently relatively small. Within this sample, all the stars on the inactive branch (marked red in Fig. \ref{fig:Rossby}) have $\rm{Ro} > 1$, while, with the exception of HD 78366, all those on the active branch (marked blue) have $\rm{Ro} < 1$. As shown in Fig \ref{fig:BVConfusogram}a, the value of $\rm{Ro} \sim 1$ also seems to separate stars with little toroidal field ($\rm{Ro} \gtrsim 1$) and those that can generate significant toroidal fields ($\rm{Ro} \lesssim 1$). It is therefore tempting to associate the active branch with toroidal fields and the inactive branch with poloidal fields. This would be a very powerful result as it would allow some information about the length of the magnetic cycle to be deduced from a measurement of the field geometry. However, we must be cautious not to over interpret the data at this stage. 

Currently, with the exception of HD 78366 ($\rm Ro = 1.2$), no active branch stars with $\rm{Ro} \gtrsim 1$ have been mapped with ZDI and hence we have little information about their field topologies. If these stars are able to generate significant toroidal fields, this would be strong evidence in favour of our hypothesis. However, if these stars turn out to be dominantly poloidal, we would need to reconsider the interpretation of the data. It is therefore important to map the surface fields of active branch stars with $\rm{Ro} \gtrsim 1$ using ZDI over their entire cycle. Within the sample of \citet{Saar1999}, there are a number of stars with $\rm{Ro} \gtrsim 1$ that possess cycle periods on the active branch, e.g. HD 165341A \& HD 190406. Under our proposed interpretation, we would expect these stars to show large $f_{\rm{tor}}$ variations over their activity cycle despite having $\rm{Ro} \gtrsim 1$. As discussed in section \ref{subsec:MagneticCycles}, HD 190406 also has a relatively short cycle period making it even more attractive as an observational target.

Similarly, in Fig. \ref{fig:BVConfusogram}b, we see that, for our sample, the two branches are almost entirely segregated by rotation period with the transition occurring at a rotation period of roughly 15 days. \citet{Petit2008} have already shown that rotation period is an important parameter determining the toroidal energy fraction. This raises the question - do stars capable of generating large $f_{\rm tor}$ values only appear on the active branch because these are the fastest rotators or is there something physically significant about the dynamos of active branch stars such that they are capable of generating large toroidal energy fractions in their surface fields? A method of breaking this degeneracy would be to map the fields of stars in the intermediate shaded regime where the branches overlap. If our hypothesis is correct, one would expect active branch stars in this region to display large toroidal energy fractions while inactive branch stars with similar rotation periods would display only poloidal fields. In table \ref{tab:ZDITargets}, we list a set of ZDI targets that would be help break the degeneracy discussed in this section. These stars all lie on the active branch in the intermediate region of Fig. \ref{fig:BVConfusogram}b. Under our interpretation, we would therefore expect them to be capable of generating strong toroidal fields. Looking at their masses and rotation periods, we find that, with the exception of HD 114710, they all lie close to, or below the $\rm Ro = 1$ curve in Fig. \ref{fig:BVConfusogram}a. This suggests that they should indeed be able to generate strong toroidal fields. It is clear that more ZDI maps and activity cycle period determinations, especially determinations of true magnetic cycle periods, will be needed before our hypothesis can be confirmed or rejected.

\section{Conclusions}
\label{sec:Conclusions}
Progress can be made in understanding stellar activity cycles by studying them in tandem with large-scale stellar magnetic field characteristics. In this paper, we have studied a sample of stars that have both (a) their large-scale magnetic fields reconstructed with Zeeman-Doppler imaging and (b) a chromospheric cycle period determination in the literature. We propose that active branch stars are able to maintain significant toroidal energy fractions with large epoch to epoch variations over the course of their activity cycle while stars that lie solely on the inactive branch remain dominantly poloidal. The reason for this behaviour may be due to different dynamo modes operating along the active and inactive branches as proposed by \citet{BohmVitense2007}. If this is indeed the case, it could provide a way to determine which branch a cycling star lies on, and hence a method of estimating a cycle period, before a cycle period determination is made. 

Despite the progress made, there are still outstanding questions. For example, why are discontinuous branches observed in the context of cycle periods but not in the activity-rotation relation? Both are manifestations of the underlying dynamo so one might naively expect them to follow similar behaviours. Possible explanations include the presence of an additional intermediate branch between the active and inactive branches \citep{Nascimento2015} or that the gap between the branches is not as distinct as currently thought (Boro-Saikia et al., in prep). Any forthcoming answers will most likely be found via theoretical simulation informed by observable constraints.

\section*{Acknowledgements}
The authors would like to thank Steven Saar for refereeing our manuscript and providing constructive criticism that helped improve this work. VS acknowledges the support of an Science \& Technology Facilities Council (STFC) studentship. AAV acknowledges support from the Swiss National Science Foundation through an Ambizione Fellowship. SBS and SVJ acknowledge research funding by the Deutsche Forchungsgemeinschaft (DFG) under grant SFB, project A16. SGG acknowledges support from the STFC via an Ernest Rutherford Fellowship [ST/J003255/1]. This study was supported by the grant ANR 2011 Blanc SIMI5-6 020 01 ``Toupies: Towards understanding the spin evolution of stars" (\url{http://ipag.osug.fr/Anr_Toupies/}). This work is based on observations obtained with ESPaDOnS at the CFHT and with NARVAL at the TBL. CFHT/ESPaDOnS are operated by the National Research Council of Canada, the Institut National des Sciences de l'Univers of the Centre National de la Recherche Scientifique (INSU/CNRS) of France and the University of Hawaii, while TBL/NARVAL are operated by INSU/CNRS.

\bibliographystyle{mnras}
\bibliography{CyclesPaper2}{}

\end{document}